\pdfoutput=1
\documentclass[cernpreprint, atlasdraft=false, texlive=2023, USenglish, texmf, orcidlogo]{atlasdoc}

\usepackage{atlaspackage}
\usepackage[normalem]{ulem}
\usepackage{bm}
\usepackage{atlasbiblatex}
 
\usepackage{atlasphysics}
 
\usepackage{multirow}
\usepackage{multicol}
\usepackage{tikz}
\usepackage[compat=1.1.0]{tikz-feynman}
\usetikzlibrary{external}
\usepackage{setspace}
 
\graphicspath{{logos/}{figures/}}
 
\usepackage{ANA-EXOT-2019-29-PAPER-defs}

\AtlasTitle{Search for heavy neutral leptons in decays of $W$ bosons using a dilepton displaced vertex in $\sqrt{s}=13$~TeV $pp$ collisions with the ATLAS detector}
 
\AtlasJournal{PRL}
\AtlasJournalRef{Phys. Rev. Lett. 131 (2023) 061803}
\AtlasDOI{10.1103/PhysRevLett.131.061803}
 
\AtlasAbstract{
 
A search for a long-lived, heavy neutral lepton ($\mathcal{N}$)
in 139~fb$^{-1}$ of $\sqrt{s}=13$~TeV $pp$ collision data collected by the ATLAS detector at the Large Hadron Collider is reported.
The $\mathcal{N}$ is produced via $W \rightarrow \mathcal{N} \mu$ or $W \rightarrow \mathcal{N} e$ and decays into two charged leptons and a neutrino, forming a displaced vertex.
The $\mathcal{N}$ mass is used to discriminate between signal and background.
No signal is observed, and limits are set on the squared mixing parameters of the $\mathcal{N}$ with the left-handed neutrino states for the $\mathcal{N}$ mass range $3$~GeV~$< m_{\mathcal{N}} < 15$~GeV.
For the first time, limits are given for both single-flavor and
multiflavor mixing scenarios motivated by neutrino flavor oscillation results for both the normal and inverted neutrino-mass hierarchies.}

\author{The ATLAS Collaboration}
 
\AtlasRefCode{EXOT-2019-29}

\PreprintIdNumber{CERN-EP-2022-017}

\hypersetup{pdftitle={ATLAS document},pdfauthor={The ATLAS Collaboration}}
 
\begin{document}
 
\maketitle

The observations of neutrino flavor oscillations~\cite{deSalas:2017kay,ParticleDataGroup:2020ssz} can be explained by postulating the existence of right-handed neutrino states that carry no Standard Model (SM) gauge charges, allowing them to have Majorana masses.
The resulting ``type-I seesaw'' model~\cite{Minkowski:1977sc,Yanagida:1979as,Glashow:1979nm,Gell-Mann:1979vob,Mohapatra:1979ia,Schechter:1980gr,Schechter:1981cv} explains the light neutrino masses and predicts heavy mass eigenstates, referred to as ``heavy neutral leptons'' (HNLs) and denoted by $\hnl$ henceforth.
The existence of HNLs can also explain the baryon asymmetry of the universe via leptogenesis~\cite{Davidson:2008bu,Pilaftsis:2009pk,Shaposhnikov:2009zzb}, which is efficient for HNL masses down to the sub-\GeV\ range~\cite{Akhmedov:1998qx,Asaka:2005an,Drewes:2017zyw,Klaric:2020phc, Klaric:2021cpi}.
Moreover, a model with three HNLs can incorporate a dark matter candidate~\cite{Asaka:2005an,Asaka:2005pn,Boyarsky:2009ix,Boyarsky:2018tvu,Ghiglieri:2020ulj}.
 
Each HNL state carries a small admixture of the left-handed neutrino of flavor $\alpha = \{e,\mu,\tau\}$.
It can therefore participate in weak interactions, controlled by dimensionless mixing coefficients $\Ua$, where $|\Ua|\ll 1$.
Previous searches were interpreted only in terms of a one-HNL model with single-flavor mixing (1SFH)~\cite{EXOT-2017-26,Belle:2013ytx,NuTeV:1999kej,DELPHI:1996qcc,CHARMII:1994jjr,NA3:1986ahv,CHARM:1985nku,WA66:1985mfx,LHCb:2014osd,LHCb:2020wxx,CMS:2022fut}.
This model is a useful benchmark but is not phenomenologically viable as it predicts neutrino masses that are too large and does not account for two neutrino mass splittings or neutrino flavor oscillations~\cite{Shaposhnikov:2006nn,Kersten:2007vk,Tastet:2021vwp}.
The simplest viable model is that of two quasi-degenerate HNLs (2QDH), with close masses and couplings, where all $\Ua$ are  nonzero.
A reinterpretation of ATLAS HNL searches in such HNL scenarios has been performed~\cite{Tastet:2021vwp}.
However, no experiment has directly explored 2QDH models yet.
 
The search reported here considers the production of HNLs via $W\to \hnl \ell_\alpha$, where  $\alpha=\{e,\mu\}$ indicates the flavor of the ``prompt'' lepton $\ell_\alpha$.
The HNL decays into two oppositely charged leptons and a neutrino: $\hnl\to \ell_\beta \ell_\gamma \nu_\gamma$ via an intermediate $W^*$ boson, or $\hnl\to \nu_\beta \ell_\gamma \ell_\gamma$ via a $Z^*$ boson, where $\beta, \gamma=e$ or $\mu$, as shown in \cref{fig:FD_LNC} (the lepton-number-violating processes are shown in Fig. 1 of the Supplemental Material~\cite{ref:supp}).
 
\begin{figure}[htb]
\hspace*{\fill}
\subfloat[ ]{
\includegraphics{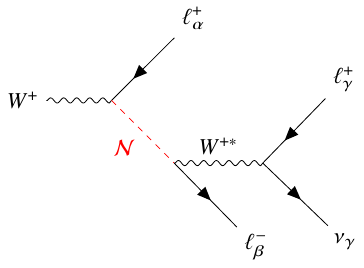}
}
\hspace*{\fill}
\subfloat[ ]{
\includegraphics{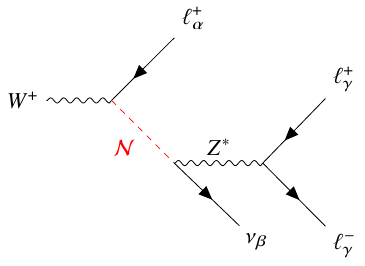}
}
\hspace*{\fill}
 
\caption{Feynman diagrams for the HNL production and decay modes targeted in this analysis. Only lepton-number-conserving processes are shown. The flavors of the leptons in the diagrams, labeled by $\alpha$, $\beta$, and $\gamma$, are either muons or electrons. If the charged leptons in the HNL decay have the same flavor, then both the diagrams with the virtual $W$ (a) and virtual $Z$ (b) contribute to the process.
Equivalent processes are also valid for an initial state $W^-$ boson.}\label{fig:FD_LNC}
\end{figure}
 
The search focuses on the mixing and mass range (up to 20~\GeV) in which the HNL is long-lived.
The resulting HNL lifetime can be approximated by $\tau_\hnl \approx (4.3\times 10^{-12}~\mathrm{s})|U|^{-2} ({m_\hnl}/1~\GeV)^{-5}$~\cite{Gronau:1984ct} where $|U|^2\equiv\sum_\beta|U_\beta|^2$, is taken from Ref.~\cite{Bondarenko:2018ptm}.
The HNL decay occurs at a significantly displaced position from the proton--proton ($pp$) collision point, forming a displaced vertex (DV) of two charged leptons, $\ell_\beta \ell_\gamma$ or $\ell_\gamma \ell_\gamma$.
The measured final states are labeled according to the prompt and displaced charged leptons therein, denoted by ``$\ell_\alpha$--$\ell_\beta\ell_\gamma$'' (explicitly listed in Table~\ref{tab:yields}).
Decays of the $W$ or \hnl~to $\tau$-leptons were determined to have negligible contribution to the analysis, since the leptonic branching fractions of the $\tau$ and the soft lepton spectrum make their selection highly inefficient.
In 1SFH scenarios, the analysis is sensitive to the squared mixing parameter \Umsq via the final states \uuu, \uue, and \uee, while  \Uesq is accessible via \eee, \eeu, and \euu.
In 2QDH scenarios, the combination of the six final states provides sensitivity to \Uesq, \Umsq, and \Utsq.
For both scenarios, bounds on the mixing parameters are extracted in the ``Dirac limit'' of lepton-number-conserving (LNC) HNL interactions, where the $W^*$-mediated final state is $\ell_\alpha^\pm$--$\ell_\alpha^\mp \ell_\gamma^\pm$, and in the ``Majorana limit'' of equal branching fractions for LNC and lepton-number-violating (LNV, $\ell_\alpha^\pm$--$\ell_\beta^\pm \ell_\gamma^\mp$) decays~\cite{ref:supp}.
The analysis can separate LNC and LNV decays only by using an explicit charge requirement for the 1SFH model in the \uue and \eeu channels, where the displaced leptons are experimentally distinguishable by their different flavors.
The bounds are tighter than and supersede those of Ref.~\cite{EXOT-2017-26}, where only the final states \uuu and \uue were studied.

This search is performed with 139 fb$^{-1}$ of 13~\TeV\ $pp$ collision data collected by the ATLAS experiment at the LHC from 2015 to 2018.
To study the signal sensitivity, Monte Carlo (MC) signal samples were generated using \Pythia[8.212]~\cite{Sjostrand:2014zea} with the A14 set of tuned parameters~\cite{ATL-PHYS-PUB-2014-021} and the \NNPDF[2.3lo] PDF set~\cite{Ball:2012cx}.
The impact of multiple $pp$ interactions per bunch crossing was modeled by adding simulated minimum-bias events generated with \Pythia[8.210] using the A3 tune~\cite{ATL-PHYS-PUB-2016-017} and \NNPDF[2.3lo] PDF set.
Particles were propagated through a detector simulation~\cite{SOFT-2010-01} based on \GEANT~\cite{Agostinelli:2002hh}.
To properly simulate spin correlations between $W$-boson decay products~\cite{Tastet:2019nqj,Ruiz:2020cjx,Tastet:2021vwp},
which are not accounted for in \Pythia[8], events are weighted to reproduce the angular distributions obtained with \MGNLO[2.9.3]~\cite{Alwall:2014hca}  using the \textsc{HeavyN} model~\cite{Degrande:2016aje,Pascoli:2018heg}.
The weighting procedure is validated by comparing the momentum spectra of each of the charged-lepton flavors and the neutrino between the weighted \Pythia[8] and \MGNLO samples.
For each $\ell_\alpha$--$\ell_\beta\ell_\gamma$ final state, signal samples were generated with HNL masses in the range $3~\GeV < m_\hnl < 20~\GeV$ and proper decay lengths $\ctau = {1, 10, 100}$~mm.

The ATLAS detector~\cite{PERF-2007-01,ATLAS-TDR-19,PIX-2018-001} is a cylindrical detector with forward-backward symmetry and nearly 4$\pi$ solid-angle coverage.\footnote{ATLAS uses a right-handed coordinate system with its origin at the nominal IP in the center of the detector and the $z$-axis along the beam pipe.
The $x$-axis points from the IP to the center of the LHC ring, and the $y$-axis points
upward. Cylindrical coordinates ($r$, $\phi$) are used in the transverse plane, $\phi$ being the azimuthal angle around the $z$-axis.
The pseudorapidity is defined in terms of the polar angle $\theta$ as $\eta = -\ln{\tan{\theta /2}}$.
Angular distance is measured in units of \(\Delta R \equiv \sqrt{(\Delta\eta)^{2} + (\Delta\phi)^{2}}\).}
It is composed of three major subsystems: the inner detector (ID) closest to the $pp$ interaction point (IP), the electromagnetic and hadronic calorimeters, and the muon spectrometer farthest from the IP.
The ID is used to reconstruct the trajectories of charged particles (tracks) in an almost uniform 2~T magnetic field, and comprises three subsystems: pixel, silicon microstrip tracker (SCT) and transition radiation tracker.
An extensive software suite~\cite{ATL-SOFT-PUB-2021-001} is used in the reconstruction and analysis of data and MC events, in detector operations, and in the trigger and data acquisition systems of the experiment.
 
Events in the signal region (SR) of this analysis were selected with triggers~\cite{TRIG-2016-01} that require a single isolated electron~\cite{TRIG-2018-05} or muon~\cite{TRIG-2018-01} with a minimum transverse momentum (\pt) of  20--26~\GeV, depending on the lepton flavor and year.
Events passing the trigger are required by a filter algorithm to contain at least one lepton~\cite{MUON-2018-03,EGAM-2018-01} with $\pt > 28~\GeV$ and $|\eta| < 2.5$.
 
To ensure isolation of this lepton from hadronic activity, the scalar sum of the $\pt$ of other tracks within a cone of size $\Delta R = 0.3$  around the lepton momentum (\isol) is required to be less than 5\% of the lepton $\pt$.
The filter also requires at least one additional lepton  with $\pt > 5$~\GeV, $|\eta| < 2.5$, and $\isol/\pt < 1.0$.
To reduce the number of events with prompt decays while maintaining efficiency for displaced leptons, the second lepton must have a transverse impact parameter ($d_0$) with respect to the IP of $|d_0| > 0.1$~mm ($|d_0| > 1$~mm) for muons (electrons).
Events that pass the filter are then processed with a large-radius tracking (LRT) algorithm~\cite{ATL-PHYS-PUB-2017-014}, that is efficient for tracks with $|d_0| < 300$~mm.
The LRT is run using hits leftover after the standard tracking~\cite{PERF-2015-08}, which is efficient only for $|d_0| < 10$~mm.
Standard and large-radius tracks are combined with muon-spectrometer tracks  (electromagnetic energy clusters) to reconstruct muons (electrons).
Events are required to contain a reconstructed primary vertex (PV) with at least two tracks, each having $\pt > 500$~\MeV.
When more than one PV is reconstructed, the one with the highest $\Sigma\pt^2$ is used, where the sum is over the tracks associated with the PV.
 
Event selection relies on the reconstruction of two physics objects: a prompt lepton and a DV.
The prompt-lepton candidate, $\ell_\alpha$, is taken to be the highest-$\pt$ muon (electron) that satisfies $\pt > 3$ (4.5)~\GeV, $|d_0| < 3$ mm, and $|(z_0 - z_{\mathrm{PV} })\sin{\theta}| < 0.5$~mm, where $z_0$ is the track's longitudinal impact parameter and $z_{\mathrm{PV} }$ is the $z$ coordinate of the PV.
If a prompt muon and a prompt electron have an angular separation $\Delta R < 0.05$, the event is rejected.
Reconstruction of DVs is performed with an optimized version of the secondary vertexing algorithm described in Ref.~\cite{ATL-PHYS-PUB-2019-013}.
First, ``seed'' DVs are formed from pairs of tracks from both the standard tracking and LRT algorithms.
Subsequently, tracks are added to the DVs, and closely spaced DVs are merged.
The secondary vertexing is run with the following configuration changes relative to Ref.~\cite{ATL-PHYS-PUB-2019-013}:
seed DVs are formed from leptons only, with at least one lepton satisfying $|d_0| > 1$~mm, and each having at least eight pixel plus SCT hits; leptonic and hadronic tracks are subsequently attached to the DVs, but
selected DVs must have exactly two leptons and no additional tracks.
 
Events must contain a prompt lepton and a DV comprising a pair of leptons with opposite-sign (OS) electric charge, although same-sign (SS) DVs are retained and used for background studies.
If a displaced track is identified as both a muon and an electron, the track is taken to be a muon (electron) if the muon- (electron-) identification quality is stricter.
The displaced muons (electrons) must have $\pt > 3$ (4.5)~\GeV.
If a displaced track in the DV is within $\Delta R = 0.05$ of the prompt lepton, the event is rejected.
The DV radial position ($\rdv$) must satisfy $4~\mathrm{mm}  < \rdv < 300$~mm.
The invariant mass of the DV and the prompt lepton, which is generally smaller than the $W$ mass, must satisfy $40~\GeV < \mdvl<90~\GeV$.
 
Background arises from five sources: DVs from particle interactions with detector material; decays of metastable SM particles; $Z \rightarrow \ell\ell$ decays; cosmic-ray muons; and DVs from random crossings of lepton tracks.
The following SR selection is designed to retain high signal efficiency and suppress the first four types of background to negligible levels, with random-crossing remaining the dominant background.
Cosmic-ray muons, which can be reconstructed as two back-to-back muons in a DV, are rejected by requiring the two displaced tracks to satisfy $\sqrt{(\Sigma\eta)^2 + (\pi - \Delta\phi)^2} > 0.05$~\cite{SUSY-2017-04}.
Dielectron ($ee$) DVs have the most background from particle interactions with detector material, so those selected must be in regions without detector material, determined from a three-dimensional map of the ID~\cite{SUSY-2016-08}.
The displaced dilepton's invariant mass ($\mdv$), which is generally smaller than $m_\hnl$ due to the unobserved final-state neutrino, is used to suppress background from $J/\psi$ and other heavy-flavor decays.
For $\uu$ DVs, $\mdv>5.5~\GeV$ is required.
For $\eu$ and $ee$ DVs, the selection efficiency is smaller, motivating looser requirements that exploit correlations between \rdv and \mdv.
These requirements are: $\mdv>5.5~\GeV$ for $\rdv<(225/7)$~mm; $\mdv>2~\GeV$ for $\rdv>(750/7)$~mm;
and $\mdv > 7~\GeV \times (1 - \rdv/(150~\mathrm{mm}))$ between these \rdv regions~\cite{ref:supp}.
 
Background from $Z\to\ell\ell$ decays, in which one of the leptons forms a DV with a third lepton, is suppressed by vetoing events where the invariant mass of the prompt lepton and the displaced lepton with the same flavor (i.e., $\alpha=\beta$)  and opposite charge satisfies $80~\GeV < m(\ell_\alpha ^\pm \ell_\beta^\mp)<100~\GeV$.
In channels with $e\mu$ DVs, the random crossing background is reduced by roughly 50\% for 1SFH, LNC interpretations,
by requiring the prompt and displaced lepton with the same-flavor to have opposite charges
: \uueLNC or \eeuLNC.
 
\begin{figure}[!ht]
\begin{center}
\begin{tabular}{cc}
 
\subfloat[]{\includegraphics[width=\figw\textwidth]{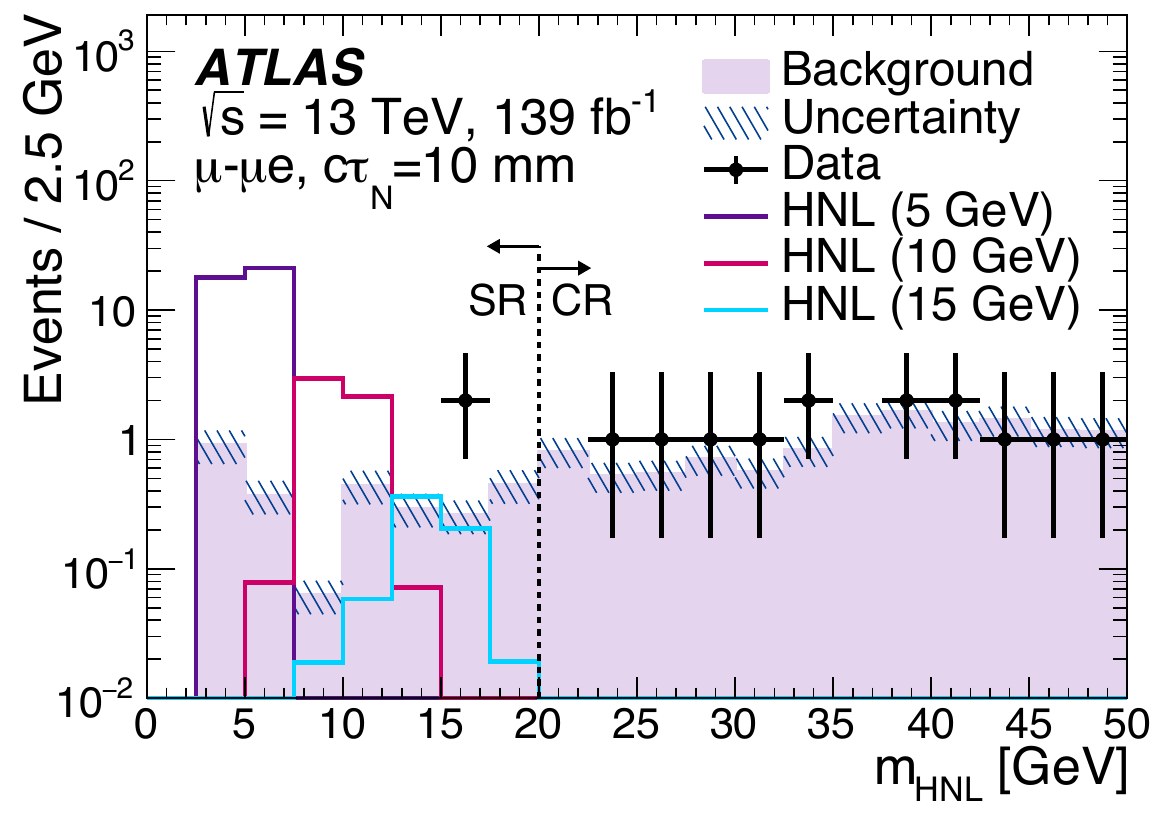}\label{fig:HNLm_uue}}
\subfloat[]{\includegraphics[width=\figw\textwidth]{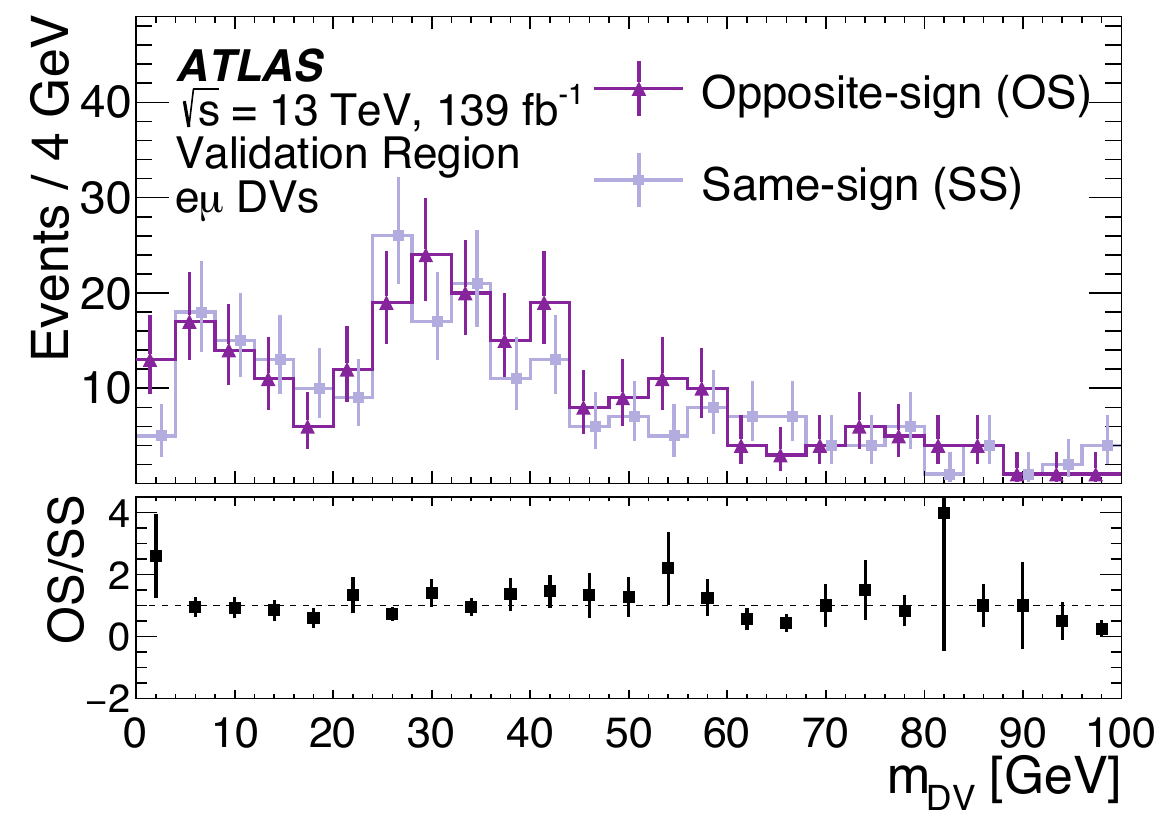}\label{fig:DV_mass_ratio}}
 
\end{tabular}
\caption{(a) The $\mhnl$ distribution in the signal (SR) and control (CR) regions for the observed data, the shuffled-event-model background normalized by the fit described in the text with its uncertainty, and simulated signal for three different mass hypotheses.
(b) The \mdv distributions for the OS and SS $e\mu$ DVs in the validation region.
The marker is offset from the central position for visualization purposes.
}
\end{center}
\end{figure}

The four-momentum of the HNL is obtained by applying four-momentum conservation in the $W$ and $\hnl$ decays, using the kinematics of the charged leptons, the known $W$ mass, an approximation where the leptons and neutrino are massless, and the flight direction of the $\hnl$, given by the vector connecting the PV and DV (see the Appendix).
This calculation yields a quadratic equation with two solutions. The positive-radical solution is used to define the invariant mass ($\mhnl$) of the HNL candidate.
In MC signal events, the distribution of $\mhnl$ peaks at the generated value $m_\hnl$, as shown in Figure~\ref{fig:HNLm_uue}.
 
The final SR selection is $\mhnl < 20~\GeV$.
The maximum signal selection efficiency is approximately 4\%.
A control region (CR) is defined as events with $20~\GeV < \mhnl < 50~\GeV$.
Since HNLs with $m_\hnl>20~\GeV$ and $|U_{\alpha}|^2$ values that the search is sensitive to are short-lived, they fail the \rdv requirements, resulting in negligible signal contamination in the CR.
 
A validation region (VR) is used for data-driven background modeling and evaluation of systematic uncertainties.
The VR comprises events that passed a variety of triggers, underwent LRT reconstruction, and do not contain a prompt lepton.
The DVs in the VR must satisfy the $\rdv$ requirements and pass the cosmic-ray muon veto.
For $ee$ DVs, the detector material veto is also applied.
The expected signal contamination in the VR is less than two events for a 100\% HNL branching fraction into the channel of interest. Since the VR contains more than 100 events in each DV channel, the signal contamination is negligible.
 
Background from random track crossings is expected to yield equal numbers of OS and SS DVs, given the large number of tracks produced in each event.
By contrast, background from $Z\to \ell\ell$ or cosmic-ray muons yields only OS DVs, and backgrounds from particle interactions with detector material or from decays of metastable hadrons preferentially yield OS DVs.
Figure~\ref{fig:DV_mass_ratio} shows the \mdv distributions for SS and OS DVs in the VR.
Good agreement is seen between the yield and shape of the distributions, shown for $e\mu$ DVs.
This indicates that the dominant source of background in the SR is random lepton crossings.
Therefore, the background model described next focuses on this background type.
A systematic uncertainty related to this assumption is described below.

\FloatBarrier
 
The signal and background yields are obtained with the following fit.
The fit uses a data-driven background model obtained from a sample of ``shuffled events''.
This sample is created by combining each OS DV in the VR with each prompt lepton found in a non-VR event that contains an SS DV satisfying loose requirements: $\mdv>1~\GeV$, with no lepton identification criteria imposed on its displaced leptons.
For each channel, the shuffled sample has at least $2\times 10^3$ times the number of events in the ``unshuffled'' data sample, in which the DV and the prompt lepton are from the same event.
As with an unshuffled event, a shuffled event may have $\mhnl<20~\GeV$ (SR) or $20~\GeV < \mhnl<50~\GeV$ (CR).
The background model in the SR and CR is given by the shuffled events (shown in Figure~\ref{fig:HNLm_uue}) with an independent floating normalization factor for each channel.
The signal model for the fit is taken from simulation and is assigned a single floating signal strength for all channels.
The input to the fit is the OS-event yields observed in the SR and CR.
Inclusion of the CR in the fit directly constrains the predicted background yield in the SR.
 
The shuffled-event background model relies on the assumption that the absence of correlation between the randomly crossing tracks results in an absence of correlation between the DV and the prompt lepton.
The validity of this assumption is checked by comparing the $\mhnl$ distributions and the $\mdvl$ distributions of shuffled events with the distributions of unshuffled events.
Only SS DVs are used in this test.
In order to have a sufficient number of unshuffled events, the requirements on \mhnl, $ m(\ell_\alpha ^\pm \ell_\beta^\mp)$, and $\mdvl$ are removed, and that on \mdv is loosened to $\mdv>2$~\GeV.
The unshuffled-event samples have between 36 and 187 events in each channel, and the shuffled-event samples are more than 50 times larger.
The comparison based on a Kolmogorov--Smirnov test yields probabilities ranging from 20\% to 99\% for the different channels, indicating the validity of the no-correlation assumption.

Systematic uncertainties in the background model, taken to be 100\% correlated between the CR and the SR, are evaluated for two sources.
The first estimates the uncertainty from the assumption that nonrandom backgrounds are negligible, and is estimated from differences between the \mhnl distributions of shuffled events created from SS and OS DVs.
This uncertainty varies between 5\% for the \eeu channel and 79\% for the \uuu channel.
The second uncertainty accounts for statistical fluctuations in the \mhnl distribution of the shuffled sample due to the finite number of prompt leptons used therein.
It is estimated from the differences between the \mhnl distributions for shuffled events of two types: in type 1 (2), the combined DV and prompt lepton originate from events in identical (different) DV channels ($ee$, \uu, \eu).
This uncertainty is largest for the \uue channel, reaching 5\%.

The total systematic uncertainty of the signal efficiency varies between 8\% and 42\% depending on the channel, $m_\hnl$, and $\ctau$.
Its largest contribution (up to 28\%) arises from the reconstruction of displaced tracks and vertices.
This is evaluated by comparing $\KzeroS\to\pi^+\pi^-$ event yields in the VR with those in MC samples produced with \Pythia[8.186] in bins of \pt and \rdv, as in Ref.~\cite{EXOT-2018-57}.
An additional uncertainty of 3\% in the track reconstruction efficiency is calculated by randomly removing tracks from each signal MC event with a \pt- and $\eta$-dependent probability~\cite{ATL-PHYS-PUB-2015-051}.
 
Uncertainties due to data--MC differences in the trigger efficiency~\cite{TRIG-2018-05,TRIG-2018-01} range up to 1\%, and those due to lepton reconstruction, identification, and impact parameter resolution are between 2\% and 17\%~\cite{EGAM-2018-01,PERF-2015-10} for the different channels.
As in Ref.~\cite{SUSY-2018-14}, an uncertainty in lepton-identification  is estimated as the difference in selection efficiency between large and small $|d_0|$ tracks. Its maximal value is 7\%.
The uncertainty in the $W$-boson production cross section and modeling is 3\%~\cite{STDM-2015-03}, and that in the HNL branching fractions and decay modeling is 5\%, arising mainly from the QCD corrections to the HNL hadronic decay width~\cite{Bondarenko:2018ptm,Davier:2005xq}.
Other uncertainties, including the impact of pileup on signal selection, luminosity uncertainty \cite{ATLAS-CONF-2019-021,LUCID2}, and uncertainty from the filtering selection used for the extended track reconstruction, each contribute at $<3$\%.
 
Table~\ref{tab:yields} shows the post-fit estimated and observed yields in the SR and CR for all channels (including the 1SFH, LNC scenario with the requirement $\ell_\alpha^\pm$--$\ell_\alpha^\mp \ell_\gamma^\pm$); a signal plus background hypothesis is used (post-fit signal is compatible with zero).
The SR contains two OS events in each of the \eee, \uue, and \uuu channels and one OS event in each of the \uee and \euu channels.
No OS \eeu events are observed.
These yields are consistent with the estimated backgrounds shown.
The observed yields in the CR are consistent with the CR background estimates.
 
\begin{table}[htbp]
\centering
\caption{
Numbers (yields) of estimated post-fit background events and of observed events in the signal and control regions.
The background yields shown are from the 2QDH, inverted-hierarchy, Majorana-limit fit described in the text, and include both systematic and statistical uncertainties.
The observed yields are shown for all final states. The last two rows show the 1SFH Dirac-limit, LNC configuration $\ell_\alpha^\pm$--$\ell_\alpha^\mp \ell_\gamma^\pm$.
\label{tab:yields}
}
\begin{tabular}{|c|c|c|c|c|}
\hline
Channel  & \multicolumn{2}{c|}{Signal region}
& \multicolumn{2}{c|}{Control region} \\
\cline{2-5}
& Background  & Observed
& Background  & Observed \\
\hline \hline
\eee & $0.4 \pm 0.3$ & $2$
& $3.6\pm 1.8$   & $2$    \\
\uee & $0.2 \pm 0.1$  & $1$
& $1.8 \pm 1.3$  & $1$ \\
\eeu & $0.9 \pm 0.4$  & $0$
& $4.1 \pm 1.9$  & $5$   \\
\uue & $2.8 \pm 0.8$  & $2$
& $12.2 \pm 3.2$~~ & $13$~~ \\
\euu & $1.2 \pm 0.9$  & $1$
& $2.8 \pm 1.6$  & $3$  \\
\uuu & $2.2 \pm 1.4$ & $2$
& $8.7 \pm 2.9$  &  $9$  \\  \hline
$e^\pm$--$e^\mp \mu^\pm$ & $0.6 \pm 0.3$  & $0$
& $2.4 \pm 1.4$  & $3$      \\
$\mu^\pm$--$\mu^\mp e^\pm$ & $1.9 \pm 0.6$  & $0$
& $8.1 \pm 2.6$  & 10~~     \\

\hline
\end{tabular}
 
\end{table}

Limits are set at 95\% confidence level (CL) on $|\Ua|^2$ vs.\ $m_\hnl$ for each HNL scenario, using the CL$_\mathrm{s}$ prescription~\cite{Read:2002hq} implemented in TRExFitter~\cite{Moneta:2010pm,Verkerke:2003ir,Cranmer:1456844}.
All systematic uncertainties are included in the fit by using nuisance parameters, whose post-fit values do not show any significant pull or constraint.
Each MC signal sample corresponds to specific values of $|U_\alpha|^2$ vs.\ $m_\hnl$, for which the efficiency is evaluated and a hypothesis test is performed with $10^4$ pseudoexperiments.
 
\begin{figure}[htbp]
\begin{center}
 
\includegraphics[width=0.5\textwidth]{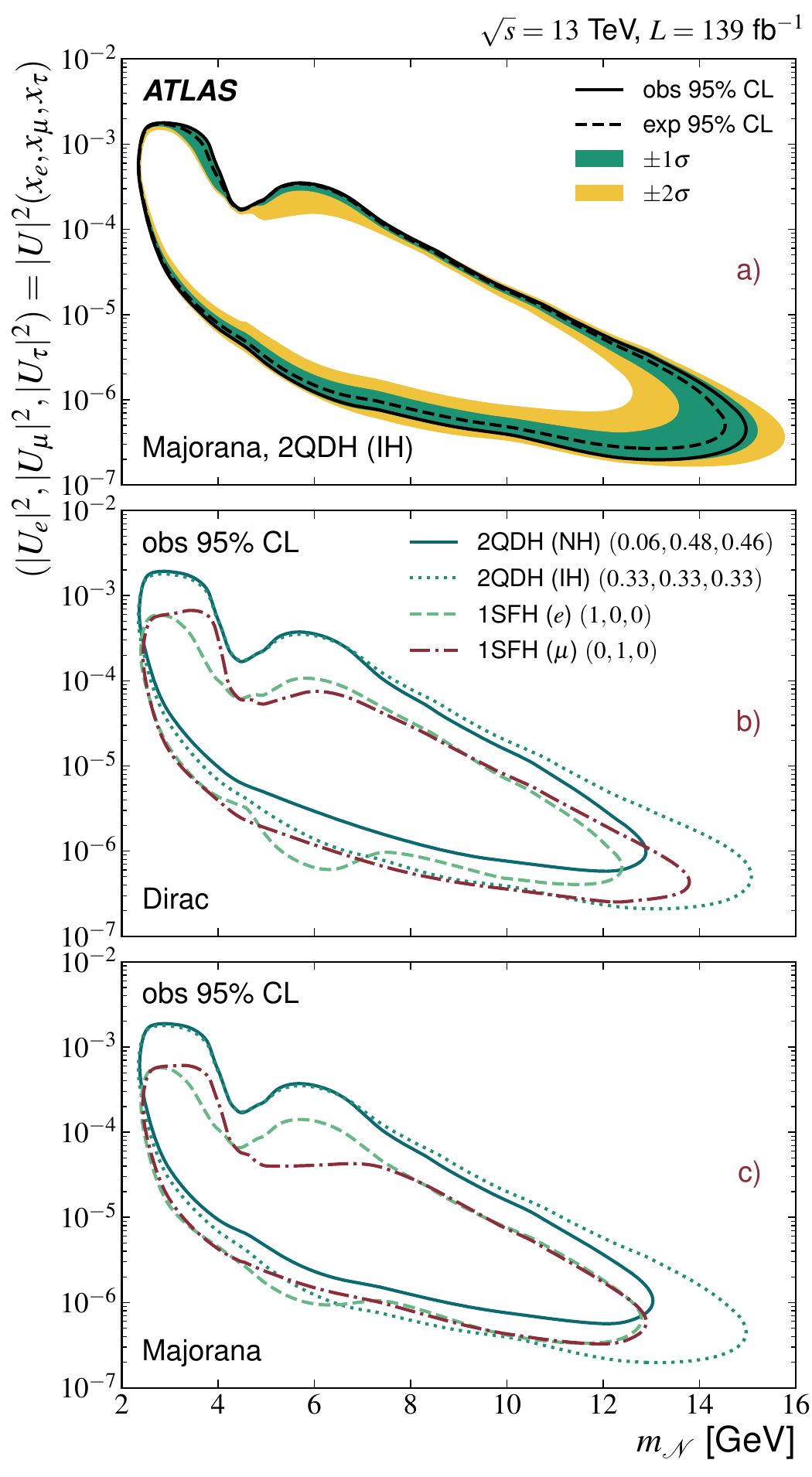}
\end{center}
\caption{(a) The observed and expected 95\% CL limits on $|\Ua|^2$ vs.\ $m_\hnl$ in the Majorana-limit case, with green and yellow bands showing the one and two standard deviation ($\sigma$) spreads for the expected limits.
(b,c) The observed limits in the 2QDH scenario with inverted (IH) and normal (NH) mass hierarchy, and in 1SFH scenarios where the HNL mixes with only $\nu_\mu$ or $\nu_e$.
}
\label{fig:limits}
\end{figure}

Figure~\ref{fig:limits} shows the excluded parameter space in the 1SFH and 2QDH scenarios for both the Dirac limit and the Majorana limit.
In the 2QDH scenarios, exclusion limits are shown for the two neutrino-mass hierarchy scenarios.
In the inverted-hierarchy case, the relative mixing coefficients are taken to be $x_\alpha \equiv |\Ua|^2/|U|^2 = 1/3$ ($\alpha=e,\mu,\tau$); for the normal-hierarchy case, the values $x_e=0.06$, $x_\mu=0.48$ and $x_\tau=0.46$ are used~\cite{Agrawal:2021dbo,Tastet:2021vwp}.
These values are at the centers of the regions consistent with the neutrino flavor oscillation data.
The observed limits are consistent with the expected limits.
The feature visible near $m_{\hnl} = 5$~\GeV~is due to the \rdv-dependent \mdv selection, which limits the sensitivity at low mass.

\FloatBarrier
 
In conclusion, a search for long-lived heavy neutral leptons is conducted in a $139~\mathrm{fb}^{-1}$ data sample of $\sqrt{s}=13~\TeV$ $pp$ collisions collected with the ATLAS detector at the LHC.
No excess is observed, and limits are set at 95\% CL on the squared mixing coefficient $|\Ua|^2$ in different HNL scenarios for HNL masses in the approximate range $3~\GeV < m_\hnl<15~\GeV$.
The observed limits exclude a region with wider ranges of $\Umsq$ and $m_\hnl$ than previously excluded by ATLAS, and the limits on $\Uesq$ are novel in ATLAS.
For the first time, limits are evaluated for the case of multiflavor mixing scenarios that agree with the neutrino flavor oscillation data, for both the normal and inverted neutrino-mass hierarchies.
The strongest limits are observed for multiflavor mixing with the inverted hierarchy.

\section*{Acknowledgments}
We wish to acknowledge our late colleague, Philippe Mermod (1978--2020), for his efforts in pioneering this search and his widespread engagement in searches for long-lived feebly interacting particles beyond the ATLAS experiment.
The search we present here could not be what it is without his vision and hard work.

% The next lines are included from the .//acknowledgements/Acknowledgements.tex input file

We thank CERN for the very successful operation of the LHC, as well as the
support staff from our institutions without whom ATLAS could not be
operated efficiently.
 
We acknowledge the support of
ANPCyT, Argentina;
YerPhI, Armenia;
ARC, Australia;
BMWFW and FWF, Austria;
ANAS, Azerbaijan;
CNPq and FAPESP, Brazil;
NSERC, NRC and CFI, Canada;
CERN;
ANID, Chile;
CAS, MOST and NSFC, China;
Minciencias, Colombia;
MEYS CR, Czech Republic;
DNRF and DNSRC, Denmark;
IN2P3-CNRS and CEA-DRF/IRFU, France;
SRNSFG, Georgia;
BMBF, HGF and MPG, Germany;
GSRI, Greece;
RGC and Hong Kong SAR, China;
ISF and Benoziyo Center, Israel;
INFN, Italy;
MEXT and JSPS, Japan;
CNRST, Morocco;
NWO, Netherlands;
RCN, Norway;
MEiN, Poland;
FCT, Portugal;
MNE/IFA, Romania;
MESTD, Serbia;
MSSR, Slovakia;
ARRS and MIZ\v{S}, Slovenia;
DSI/NRF, South Africa;
MICINN, Spain;
SRC and Wallenberg Foundation, Sweden;
SERI, SNSF and Cantons of Bern and Geneva, Switzerland;
MOST, Taiwan;
TENMAK, T\"urkiye;
STFC, United Kingdom;
DOE and NSF, United States of America.
In addition, individual groups and members have received support from
BCKDF, CANARIE, Compute Canada and CRC, Canada;
PRIMUS 21/SCI/017 and UNCE SCI/013, Czech Republic;
COST, ERC, ERDF, Horizon 2020, ICSC-NextGenerationEU and Marie Sk{\l}odowska-Curie Actions, European Union;
Investissements d'Avenir Labex, Investissements d'Avenir Idex and ANR, France;
DFG and AvH Foundation, Germany;
Herakleitos, Thales and Aristeia programmes co-financed by EU-ESF and the Greek NSRF, Greece;
BSF-NSF and MINERVA, Israel;
Norwegian Financial Mechanism 2014-2021, Norway;
NCN and NAWA, Poland;
La Caixa Banking Foundation, CERCA Programme Generalitat de Catalunya and PROMETEO and GenT Programmes Generalitat Valenciana, Spain;
G\"{o}ran Gustafssons Stiftelse, Sweden;
The Royal Society and Leverhulme Trust, United Kingdom.
 
The crucial computing support from all WLCG partners is acknowledged gratefully, in particular from CERN, the ATLAS Tier-1 facilities at TRIUMF (Canada), NDGF (Denmark, Norway, Sweden), CC-IN2P3 (France), KIT/GridKA (Germany), INFN-CNAF (Italy), NL-T1 (Netherlands), PIC (Spain), ASGC (Taiwan), RAL (UK) and BNL (USA), the Tier-2 facilities worldwide and large non-WLCG resource providers. Major contributors of computing resources are listed in Ref.~\cite{ATL-SOFT-PUB-2023-001}.

% End of text imported from the .//acknowledgements/Acknowledgements.tex input file

\FloatBarrier
 
\clearpage
\appendix

% The next lines are included from the .//HNL_mass_SM.tex input file
{\textit{Appendix: A HNL mass.}}---The HNL mass
(\mhnl) can be obtained using energy--momentum conservation in the HNL production ($W \rightarrow \mathcal{N}\ell_1$) and decay ($\mathcal{N} \rightarrow \ell_2 \ell_3 \nu$), where $\ell_1$ is the prompt lepton and $\ell_2$ and $\ell_3$ are the charged leptons in the DV.
The problem can be summarized with the following equations. Four-momentum conservation in the $\mathcal{N}$ decay gives
\begin{align}
p^\mu_{\mathcal{N}}&=p^\mu_{2}+p^\mu_{3}+p^\mu_{ \nu } \equiv p^\mu_{23}+p^\mu_{\nu}.\label{eq:HNL4vec}
\end{align}
Four-momentum conservation in the $W$ decay gives
\begin{align}
p^\mu_{W}&=p^\mu_{1}+p^\mu_{\mathcal{N}} =  p^\mu_{1} + p^\mu_{23} +p^\mu_{ \nu }.\label{eq:W4vec}
\end{align}
The following are defined
\begin{align}
p_{23}^{2}&=E_{23}^{2}- \vert \vec{p}_{23} \vert ^{2} \equiv m_{23}^{2} \nonumber \\
p_{23}^{\parallel} &\equiv \vec{p}_{23}\cdot\bm{\hat{v}} \nonumber \\
p_{23}^{\bot} &\equiv  \vert \vec{p}_{23}-p_{23}^{\parallel}\bm{\hat{v}} \vert \nonumber
\end{align}
\noindent where $m$, $E$, and $|\vec p|$ are the mass, energy, and momentum-vector magnitude of the particles indicated by their subscript and $\bm{\hat{v}}$ is the flight direction of the HNL given by the vector connecting the PV and DV.
 
The solution to the HNL mass is presented in the coordinate system $k = (\bm{\hat{x'}}, \bm{\hat{y'}}, \bm{\hat{z'}})$, which is rotated relative to the ATLAS coordinate system, such that the origin of the $k$-frame is at the PV and the $z'$-axis points along the flight direction of the HNL. The definition of this coordinate system is
\begin{align}
\bm{\hat{z'}} = \bm{\hat{v}}, \ \ \ \bm{\hat{x'}} = \frac{\vec{p}_{23} \times \bm{\hat{z'}}}{|\vec{p}_{23} \times \bm{\hat{z'}}|}, \ \ \ \bm{\hat{y'}} = \bm{\hat{z'}} \times \bm{\hat{x'}}. \nonumber
\end{align}
\noindent The momenta of $\ell_2$ and $\ell_3$ constrain the components of the neutrino momentum orthogonal to $\vec{p}_{\mathcal{N}}$. This means that energy--momentum conservation in the $W$ and $\mathcal{N}$ decays can be expressed in terms of one unknown variable $\alpha$, which is the component of neutrino momentum in the $\bm{\hat{z'}}$ direction.
To express \cref{eq:W4vec,eq:HNL4vec} in terms of $\alpha$, the following quantities are defined
\begin{align}
\vec{p'}_{23} &\equiv \vec{q} \label{eq:q_def_1} \\
\vec{q}_{\:\:} &= (0,\: |\vec{p}_{23} \times \bm{\hat{z'}}| \equiv q_{\perp},\: \vec{p}_{23} \cdot \bm{\hat{z'}} \equiv q_{z}) \label{eq:q_def_2} \\
\vec{p'}_\nu &=( 0,\: -q_{\perp}, \: \alpha)  \label{eq:q_def_3} \\
E'_\nu &= \sqrt{q_{\perp}^2+\alpha^2}. \label{eq:alpha_E}
\end{align}
Squaring \cref{eq:W4vec} gives
\begin{align}
p^{'2}_W  = m_W^2 = m_1^2 + m_\nu^2 + m_{23}^2  + 2p'_1 \cdot (p'_{23}+ p'_\nu) + 2 p'_{23} \cdot p'_\nu \label{eq:w_mass}
\end{align}
\noindent where
\begin{align}
p'_1 \cdot (p'_{23}+ p'_\nu) &= E'_1 (E'_{23}+ E'_\nu) - p'_{1,z}(q_z + \alpha)  \nonumber \\   
p'_{23} \cdot p'_\nu &= E'_{23}E'_\nu + q_{\perp}^2 - q_z\alpha.  \nonumber  
\end{align}
In the energy regime of interest, the charged leptons and neutrino can be treated as massless particles, such that $m_1 = m_\nu = 0$.
Rearranging \cref{eq:w_mass} to solve for $E_\nu$ gives
\begin{align}
E'_\nu = A  + B\alpha \label{eq:Enu}
\end{align}
\noindent where
\begin{align}
A = \frac{(m_W^2 - m_{23}^2)/2 - E'_1E'_{23} + p'_{1,z}q_z - q_{\perp}^2}{E'_1 + E'_{23}}, \ \ \ B = \frac{p'_{1,z} + q_z}{E'_1 + E'_{23}} \nonumber
\end{align}
Subtracting \cref{eq:Enu} from \cref{eq:alpha_E} gives the following quadratic expression in $\alpha$
\begin{align}
(B^2 -1)\alpha^2 + 2&AB\alpha + A^2 - q_{\perp}^2 = 0.   \nonumber  
\end{align}
\noindent The solution for $\alpha$ is therefore
\begin{align}
\alpha= \frac{-AB \pm \sqrt{(B^2-1)q_{\perp}^2 + A^2}}{(B^2-1)}. \label{eq:alpha_soln}
\end{align}
Both solutions for $\alpha$ were studied using simulated HNL events and it was noted that the solution that led to a smaller $|\vec{p}_\mathcal{N}|$ typically led to a value for \mhnl that was closer to the simulated $m_\mathcal{N}$. This solution often corresponded to forward emission of the neutrino with respect to the HNL decay. Therefore, the definition of \mhnl in the analysis uses the solution with the positive radical.
 
The expression for $\alpha$ in \cref{eq:alpha_soln} depends on $m_W$.
ATLAS has measured the $W$-boson pole mass to be $M_W=80.370 \pm 0.019$~\GeV\ \cite{STDM-2014-18}. This measurement is combined in Ref.~\cite{ParticleDataGroup:2020ssz} with results from other collider experiments to provide a measurement of the $W$-boson width, $\Gamma_W = 2.195 \pm 0.083$~\GeV.
Since the $W$ mass has a width, then if $m_W= M_W$ in \cref{eq:alpha_soln} it is possible that there is no real solution for $\alpha$.
Instead of rejecting these events, $m_W$ is set equal to the median $W$ mass in the kinematically allowed region ($m_{W, \mathrm{med}}$). This ensures that $\alpha$ (and correspondingly \mhnl) always has a real solution.
 
To define the kinematically allowed region, the minimum $W$ mass that is consistent with the charged-lepton decay products ($m_{W, \mathrm{min}}$) is computed. From \cref{eq:w_mass}, the mass of the $W$ boson is given by
\begin{align}
m_W^2 &=  m_{23}^2 +2\left(E'_1E'_{23} + E'_\nu(E'_1 + E'_{23}) - p'_{1,z}q_z + q_{\perp}^2 -\alpha(p'_{1,z} + q_z) \right) \label{eq:wmass}
\end{align}
\noindent and $m_{W, \mathrm{min}}$ occurs where
\begin{align}
\frac{d(m_W^2/2)}{d\alpha} &=  (E'_1 + E'_{23})\frac{dE'_\nu}{d\alpha} - (p'_{1,z} + q_z) = 0. \label{eq:min_W_alpha}
\end{align}
\noindent Using
\begin{align}
\frac{dE'_\nu}{d\alpha} = \frac{d\sqrt{q_{\perp}^2+\alpha^2}}{d\alpha} = \frac{\alpha}{E'_\nu} \nonumber
\end{align}
\noindent in \cref{eq:min_W_alpha}, the chosen value of $\alpha$ that gives the minimum $m_W$ is
\begin{align}
\alpha =  \frac{q_\perp B }{\sqrt{1-B^2}}. \label{eq:min_alpha}
\end{align}
Substituting \cref{eq:min_alpha} into \cref{eq:wmass}, the minimum $W$ boson mass is
\begin{align}
m^2_{W,\mathrm{min}} &=  m_{23}^2 +2\left(E'_1E'_{23}  + (E'_1+E'_{23})\sqrt{q_{\perp}^2 + \frac{q_\perp^2 B^2 }{1-B^2}  } -p'_{1,z}q_z + q_{\perp}^2- (p'_{1,z} + q_z )\frac{q_\perp B }{\sqrt{1-B^2}}\right).  \nonumber  
\end{align}
 
The cumulative probability for the $W$ boson to have a mass greater than $m_{W,\mathrm{min}}$ is used to find the median of the remaining distribution.
The probability density function ($f$) for $m_W^2$ satisfies
\begin{equation*}
f(m_W^2) \propto \frac{1}{(m_W^2-M_W^2)^2 + M_W^2\Gamma_W^2}.
\end{equation*}
Therefore, the cumulative distribution function ($F$) is
\begin{equation}
F(m_W^2) = \frac{1}{\pi} \arctan\left( \frac{m_W^2-M_W^2}{M_W\Gamma_W}\right) + \frac{1}{2}. \label{eq:cdf}
\end{equation}
The midpoint of the allowed kinematic region has a value of
\begin{equation*}
F_{\mathrm{med}} = \frac{1+ F(m_{W,\mathrm{min}}^2)}{2}
\end{equation*}
Rearranging \cref{eq:cdf} for $m_W^2$ gives
\begin{equation}
m_W^2 = M_W^2 + \Gamma_W  M_W  \tan \left(\pi\left[F-\frac{1}{2}\right]\right). \label{eq:E_cdf}
\end{equation}
Substituting $F=F_{\mathrm{med}}$ in \cref{eq:E_cdf}
gives an expression for the median $W$ mass in the kinematically allowed region
\begin{align}
m_{W, \mathrm{med}}^2 &=  M_W^2 + \Gamma_W M_W \tan \left(\pi\left[\frac{1+ F(m_{W,\mathrm{min}}^2)}{2}-\frac{1}{2}\right]\right).  \nonumber  
\end{align}
This value of $m_{W,\mathrm{med}}$ is used in \cref{eq:alpha_soln} to solve for $\alpha$.
 
From \cref{eq:HNL4vec} and the definitions in \cref{eq:q_def_1,eq:q_def_2,eq:q_def_3,eq:alpha_E}, the expression for the HNL mass in terms of $\alpha$ is
\begin{align}
\mhnl^2 &= m_{23}^2 + 2p'_\nu \cdot p'_{23} \nonumber \\
&= m_{23}^2 + 2E'_{23}\sqrt{q_{\perp}^2+\alpha^2} + 2q_{\perp}^2 - 2q_z\alpha. \label{eq:HNL_mass}
\end{align}
Substituting the expression for $\alpha$ in \cref{eq:alpha_soln} into \cref{eq:HNL_mass} gives the solution for the HNL mass.

% End of text imported from the .//HNL_mass_SM.tex input file

\printbibliography

\clearpage
% ATLAS Collaboration author list
% Reference date of EXOT-2019-29 is 2021-11-02
% Author list last updated on date 30-NOV-23
% Data extracted on 30-Nov-2023 for paper reference EXOT-2019-29
% at 7:12pm
 
\begin{flushleft}
\hypersetup{urlcolor=black}
{\Large The ATLAS Collaboration}

\bigskip

\AtlasOrcid[0000-0002-6665-4934]{G.~Aad}$^\textrm{\scriptsize 101}$,
\AtlasOrcid[0000-0002-5888-2734]{B.~Abbott}$^\textrm{\scriptsize 119}$,
\AtlasOrcid[0000-0002-7248-3203]{D.C.~Abbott}$^\textrm{\scriptsize 102}$,
\AtlasOrcid[0000-0002-2788-3822]{A.~Abed~Abud}$^\textrm{\scriptsize 36}$,
\AtlasOrcid[0000-0002-1002-1652]{K.~Abeling}$^\textrm{\scriptsize 55}$,
\AtlasOrcid[0000-0002-2987-4006]{D.K.~Abhayasinghe}$^\textrm{\scriptsize 94}$,
\AtlasOrcid[0000-0002-8496-9294]{S.H.~Abidi}$^\textrm{\scriptsize 29}$,
\AtlasOrcid[0000-0002-9987-2292]{A.~Aboulhorma}$^\textrm{\scriptsize 35e}$,
\AtlasOrcid[0000-0001-5329-6640]{H.~Abramowicz}$^\textrm{\scriptsize 150}$,
\AtlasOrcid[0000-0002-1599-2896]{H.~Abreu}$^\textrm{\scriptsize 149}$
% ...
\bigskip
\\

% ...
$^{29}$Physics Department, Brookhaven National Laboratory, Upton NY; United States of America.\\
% ...
$^{35}$$^{(a)}$Facult\'e des Sciences Ain Chock, R\'eseau Universitaire de Physique des Hautes Energies - Universit\'e Hassan II, Casablanca;$^{(b)}$Facult\'{e} des Sciences, Universit\'{e} Ibn-Tofail, K\'{e}nitra;$^{(c)}$Facult\'e des Sciences Semlalia, Universit\'e Cadi Ayyad, LPHEA-Marrakech;$^{(d)}$LPMR, Facult\'e des Sciences, Universit\'e Mohamed Premier, Oujda;$^{(e)}$Facult\'e des sciences, Universit\'e Mohammed V, Rabat;$^{(f)}$Institute of Applied Physics, Mohammed VI Polytechnic University, Ben Guerir; Morocco.\\
$^{36}$CERN, Geneva; Switzerland.\\
% ...
$^{55}$II. Physikalisches Institut, Georg-August-Universit\"{a}t G\"ottingen, G\"ottingen; Germany.\\
% ...
$^{94}$Department of Physics, Royal Holloway University of London, Egham; United Kingdom.\\
% ...
$^{101}$CPPM, Aix-Marseille Universit\'e, CNRS/IN2P3, Marseille; France.\\
$^{102}$Department of Physics, University of Massachusetts, Amherst MA; United States of America.\\
% ...
$^{119}$Homer L. Dodge Department of Physics and Astronomy, University of Oklahoma, Norman OK; United States of America.\\
% ...
$^{149}$Department of Physics, Technion, Israel Institute of Technology, Haifa; Israel.\\
$^{150}$Raymond and Beverly Sackler School of Physics and Astronomy, Tel Aviv University, Tel Aviv; Israel.\\
% ...
$^{a}$ Also Affiliated with an institute covered by a cooperation agreement with CERN.\\
$^{b}$ Also at Borough of Manhattan Community College, City University of New York, New York NY; United States of America.\\
$^{c}$ Also at Bruno Kessler Foundation, Trento; Italy.\\
$^{d}$ Also at Center for High Energy Physics, Peking University; China.\\
$^{e}$ Also at Centro Studi e Ricerche Enrico Fermi; Italy.\\
$^{f}$ Also at CERN, Geneva; Switzerland.\\
% ...

\end{flushleft}

% Created with Glance <Atlas.Glance@cern.ch>

\end{document}